\documentstyle[aps,twocolumn]{revtex}
\input psfig
\begin{document}
\title{Nonlinear ac susceptibility studies of  high-$T_c$ rings: Influence of the structuring method and
determination of the flux creep exponent}
\author{S. Streubel, F. Mrowka, M. Wurlitzer, and P.  Esquinazi}
\address{Abteilung Supraleitung und Magnetismus, Institut f\"ur Experimentelle Physik II, Universit\"at
  Leipzig, Linn\'estrasse 5, D-04103~Leipzig, Germany}
\author{K. Zimmer}
\address{Institut f\"ur Oberfl\"achenmodifizierung e.V.,
 D-04138~Leipzig, Germany}
\maketitle
\begin{abstract}
We have studied the influence of the patterning procedure on the critical
current density of high-$T_c$ YBa$_2$Cu$_3$O$_{7-\delta}$ thin rings using
the nonlinear ac susceptibility method. At no applied dc magnetic field we
have found that laser ablation degrades strongly the critical current density whereas
ion beam etching has only a weak influence on it. From the measurements at different
frequencies and dc magnetic fields we analyzed the influence of flux creep and
obtained the field dependence of  the flux creep exponent. Our data reconfirm
the recently observed scaling relation for the nonlinear susceptibility response 
of type-II superconductors.
\end{abstract}
\draft 

\section{Introduction}
In a recently published letter we have shown that the measurement of the 
magnetic moment of samples with ring geometry 
enables the identification of regions with degraded superconducting regions and
can be used as a sensitive method to investigate the critical current density $J_c$ of high
temperature superconducting thin films
\cite{Mrowka97feb1}. 
 The ac field amplitude dependence of
the ac susceptibility of structured narrow rings provides the
experimental foundation for this method. Within the Bean model and via the determination 
of the so-called penetration field
$H_p$ (at which the perfect diamagnetic shielding is
lost) and geometry parameters we obtain $J_c$.  
In this work
we have 
exploited this technique at 
zero dc magnetic field $H_{dc}$
to investigate the influence of the
structuring method used to fabricate the high-temperature superconducting 
rings on $J_c$. Our results indicate clearly that ion beam etching is a far less destructive
structuring technique than laser ablation.

We have also obtained the dc field dependence of $J_c$ and analysed the
effects of flux
creep with an extended Bean model which enables the interpretation
of the frequency dependence of the ac susceptibility 
associated with a finite resistivity due to flux motion \cite{Brandt95}. We have
analyzed the shift of the ac susceptibility with frequency and
applied a scaling relation which has been  predicted by
Brandt \cite{Brandt97jun1}. Within this theory we have determined the flux
creep exponent $n(T,H_{dc})$ and compared it with the one determined
by relaxation measurements with a SQUID. 

The paper is organized as follows. In the next section we give a brief
summary of the measured samples and experimental details. In section
III we provide a theoretical resume of the ac response of narrow superconducting rings as well as
the determination of the flux creep exponent from ac susceptibility. The
results are described in section IV. A brief summary is given in section V.

\section{Experimental Details and Investigated Samples}

The ac susceptibility measurements were performed with a Lake Shore 7000
AC Susceptometer designed for operation at low-level ac magnetic
field amplitudes $H_0 \leq$ 2 mT.  We applied dc magnetic fields up to 3 T by a
superconducting solenoid.

We have studied rings made from YBa$_2$Cu$_3$O$_{7-\delta}$ high-temperature superconducting
200 nm thin films prepared
by pulsed laser deposition on 1 mm thick $\rm Al_2O_3$ substrates with
a critical temperature between \mbox{$T_c$ = 89 K} and 90 K
\cite{Lorenz94nul1,Lorenz96jun1}.  A 30 nm $\rm CeO_2$ buffer
layer was used.  All the films were characterized by ac
susceptibility measurements before structuring. Results on the films
can be found
in \cite{Wurlitzer97may1}. The films were structured by two methods:
(a) by laser ablation (LA) with an excimer workstation having an
  optical resolution of $1.5 \rm \mu m$ and
(b) by ion beam etching (IE). For the ion beam etching procedure 
a spun-on $1.5~\mu$m thick resist layer (AZ1450) was structured by 
excimer laser ablation using a scanning method. The resist was etched by 
laser apart from a thin, $\sim 200$~nm thick remaining film to prevent any damage 
of the superconducting layer. We used ion beam sputtering with argon to 
transfer the mask structure into the film. The ion etching was performed in a 
non-commercial IBE system with a base pressure better than $2 \times 10^{-6}$~mbar 
equipped with a Kaufman type ion beam source operating at a beam energy of 700~eV 
and current density of 0.2~mA/cm$^2$. The beam was neutralized by means of a 
hot filament to prevent charge accumulation on the sample surface. The samples 
were mounted on a water-cooled and rotating sample holder to avoid an excessive 
increase of temperature during  etching.
Table \ref{samples} shows the main characteristics of  the measured samples.

\section{Theory}

\subsection{AC Susceptibility of Narrow Superconducting Rings}\label{s:ac}

In this section we briefly review the ac response of narrow superconducting rings 
as described in \cite{Mrowka97feb1,Brandt97jun1}. We
discuss the limit of a thin narrow ring of width $w$ much
smaller than the mean radius $R$, \mbox{$R-w/2 \le r \le R+w/2$}. Within the Bean model the virgin
magnetization curve completely determines the hysteresis loop which
has the shape of
a parallelogram \cite{Ishida81nul1,Gilchrist93nul1}. From these loops
one obtains the complex susceptibility \mbox{$\chi = \chi' - i
  \chi''$} of a superconducting narrow ring with constant $J_c$.  For
cycled magnetic field \mbox{$H_a(t) = H_0 \sin{\omega t}$} we
define the nonlinear susceptibility, the quantity we measure, as 
\begin{equation}\label{eq:chi}
  \chi(H_0) = \frac{\omega}{V \pi H_0} \int_0^{2 \pi} m(t) e^{-i \omega
  t} dt.
\end{equation}
where $m$ is the magnetic moment and $V$ the sample volume.
The virgin magnetization curve of such a ring is composed of two straight
lines, \mbox{$m(H_a) = (H_a/H_p)m_{sat}$} for \mbox{$H_a \le H_p$} and
\mbox{$m(H_a) = m_{sat}$} for \mbox{$H_a \ge H_p$}, since 
the screening supercurrent in the ring is limited to
a maximum value \mbox{$I_c = J_c d w$} ($d$ is the ring thickness). As long
as \mbox{$|I| < I_c$} holds for
the current induced in the ring by the applied field $H_a$, 
 no magnetic flux can penetrate through the ring
into the hole of the ring. When $|I| = I_c$ is reached the ring becomes
transparent to magnetic flux. Therefore, when the applied field is
increased further, flux lines will move through the ring as described
in \cite{Brandt93jun1,Brandt93nov1}. These authors treat the
superconducting strip with transport current in an applied field. With
the moving 
flux lines in the ring, magnetic flux enters the ring hole until the
screening supercurrent decreases again to the value $I_c$. The magnetic
moment \mbox{$m = \pi R^2 I$} of the ring thus saturates at the value
\mbox{$m_{sat} = \pi R^2 I_c$}. The applied field value $H_p$ at which
this saturation is reached follows from the inductivity $L$ of the flat
narrow ring \cite{Brandt96aug1},
\mbox{$L = \mu_0 R ( \ln{8 R / w} - a)$} with $a \simeq 0.5$
\cite{Landau63nul1}.

The magnetic flux generated in the hole by a ring current $I$ is 
$\mit\Phi = L I$. As long as $I < I_c$ one has ideal screening, thus 
\mbox{$\mit\Phi = - \pi R^2 \mu_0 H_a$}. Equating these two fluxes one
obtains for the flat ring \mbox{$I = \mit\Phi / L = - \pi R H_a /
  [\ln{8 R   / w} - a]$}. At $H_a = H_p$ one reaches \mbox{$|I| =
  I_c = J_c w d$}, thus \mbox{$H_p = [\ln{8 R / w} - a]I_c /
  \pi R$}. Therefore, one can express $J_c$ as 
\begin{equation}
  \label{eq:jc}
  J_c = \frac{\pi R}{w d \Big[\ln{\frac{8 R}{w}}-
  \frac{1}{2}\Big]} H_p\,. 
\end{equation}
The
formulae given above are accurate to corrections  of order $w / 2R$
 since the precise values of $m$ and $L$ depend on
the current distribution across the ring width $w$ which changes
during the magnetization process. 

With Eq.~(\ref{eq:chi}) and the expressions for $m(t)$ which are
given in \cite{Brandt97jun1} we obtain the susceptibility of the ring normalized to the
initial value $\chi(0) = -1$, i.e., to $\chi \to \chi V / m'(0)$,
\begin{eqnarray}
\label{eq:x}
   \chi'(h) & = & -1, \quad \chi''(h) = 0, \quad h \le 1,\nonumber \\
   \chi'(h) & = & \displaystyle -\frac{1}{2} - \frac{1}{\pi} \arcsin{s} +
  \frac{1}{\pi}  s \sqrt{1 - s^2}, \nonumber \\ 
   \chi''(h) & = & \displaystyle \frac{4}{\pi} \frac{h -1}{h^2}, \quad h \ge 1,
\end{eqnarray}
with $h = H_0 / H_P$ and $s = 2/h -1$.
These theoretical results have been confirmed experimentally in a previous paper
\cite{Mrowka97feb1}. We use expression 
(\ref{eq:jc}) to determine $J_c$ from the measured penetration field $H_p$.
 
\subsection{Flux Creep Exponent from ac Susceptibility}\label{stheory}

For high enough dc magnetic fields, the influence of flux creep should be
taken into account in the ac response of the superconducting ring.
In our analysis we use a current voltage law $E=E(J)$ that is
connected to a logarithmic dependence \cite{Zeldov90feb,Vinokur91aug1}
of the pinning potential on the current density
\mbox{$U(J)=U_0\ln(J_c/J)$} and describes flux creep as an activated
motion of vortices which have to overcome this potential. Following
\cite{Vinokur91aug1} we have
\begin{equation}\label{ecvlaw}
E(J)=E_c\exp^{-\frac{U(J)}{kT}}=E_c\left(\frac{J}{J_c}\right)^{n}\,,
\end{equation}
where $n(T,H)={U_0(T,H)/kT}$ is the so-called flux creep exponent. The
power law in Eq.~(\ref{ecvlaw}) includes the limiting cases  of
Bean-like field distribution for which $n=\infty$ and of  Ohmic dissipation for which $n=1$.

By investigating the frequency ($f$) dependence of narrow rings under the
application of a harmonic ac field $H_a(t)=H_0 \sin(\omega t)$ during flux creep, 
\mbox{e.g. $1<n<\infty$}, Brandt \cite{Brandt97jun1} derived a scaling law 
for the frequency and magnetic field amplitude which states the following:
 The complex ac susceptibility at fixed temperature 
\mbox{$\chi(T,f,H)=\chi'-i \chi''$} remains unchanged under the 
simultaneous transformation of time by a constant factor 
$C$, e.g. $t \rightarrow t/C$, and magnetic field by a factor  $C^{\beta}$, 
e.g. $H \rightarrow HC^{\beta}$, with $\beta=1/(n-1)$. 
This is equivalent to state that the following equation holds:
\begin{equation}\label{etransform}
\chi\left(T,f,H \right)=\chi\left(T,fC,HC^{\beta} \right)\,.
\end{equation}
Using this property we determined the flux creep exponent
$n(T,H)$ at different $T$ and applied dc fields $H_{dc}$ by
measuring the ac susceptibility at different frequencies
\mbox{$f_0,f_1,f_2 \ldots f_i$} and by equating the ac fields at which
the same susceptibility is measured, i.e.
\begin{equation}\label{esusc1}
H_i=H_0C^{\beta}= H_0{(f_i/f_0)}^{\frac{1}{n-1}}\,,
\label{hsca}
\end{equation}
where $C$ is determined by the frequency ratio $C=f_i/f_0$. Equation (\ref{esusc1}) 
is a power law for the magnetic field amplitude. 
Plotting $\ln H$ vs. $\ln f$ should then give a straight line of slope $\beta=1/(n-1)$.\\

\begin{figure}
\centerline{\psfig{file=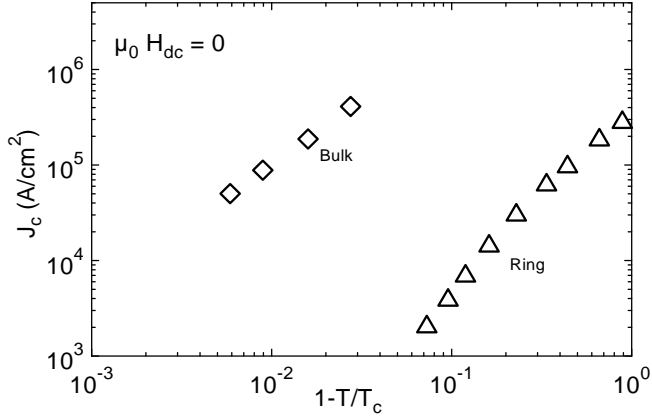,width=91mm}}
\caption{Critical current density at zero applied dc field of sample LA2
as a function of reduced temperature. Note the
strong decrease in $J_c$ after structuring by laser ablation.} 
\label{jcLA2}
\end{figure}

J{\"o}nsson-{\AA}kerman et al. ~recently confirmed the 
scaling relation for the nonlinear ac susceptibility response of
HgBa$_2$CaCu$_2$O$_{6+\delta}$ thin films \cite{Jonsson99}. In a previous work \cite{Jonsson98} 
the flux creep exponent $n$ was determined
from the frequency dependence of $\chi'$
in the limit of large ac amplitudes. In what follows we present
a further confirmation of the scaling relation (\ref{etransform})
by applying it on structured thin rings. Furthermore, we use a
straightforward analysis procedure to obtain the
field dependence of the pinning potential $U_0$ for fields up to 2T.

\section{Results}

\subsection{Influence of the structuring method on
  $J_c$}\label{sstructure}

In \cite{Mrowka97feb1} a drastic decrease of $J_c$ after
structuring the films into rings was found. This behavior has been explained
by the existence of microcracks in the path of the ring. In this paper we 
show that the structuring technology used to fabricate the superconducting ring 
has a large influence on its $J_c$.

\begin{figure}
\centerline{\psfig{file=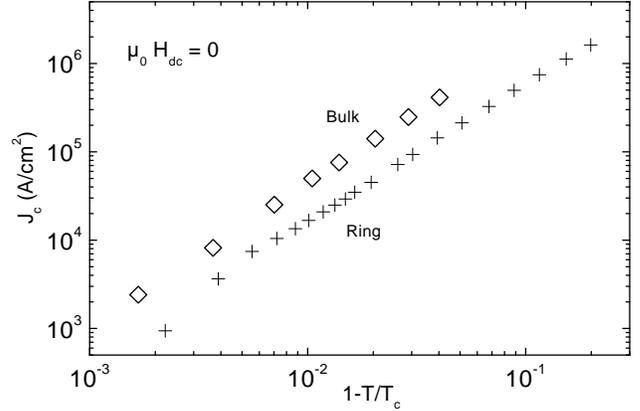,width=91mm}}
\caption{Critical current density at zero applied dc field of sample IE1
as a function of reduced temperature. The patterning of the film by
ion beam etching produces a relatively small
decrease of $J_c$.} 
\label{jcIE1}
\end{figure}
Figure \ref{jcLA2} shows the temperature dependence of the critical
current density $J_c$ for sample~LA2 before and after structuring by
laser ablation. For this sample and before structuring $J_c$ is
500 times larger than its value after structuring. For comparison
the $J_c$ values for the sample~IE1 before and after structuring by
ion beam etching are given in Fig.~\ref{jcIE1}. 
For the ring of this sample $J_c$ is reduced only  30 \% compared
to the unstructured sample.
For all measured samples we have found that
$J_c$ of the measured LA-Rings is at least 100 times smaller than
$J_c$ before structuring. On the other hand the results for the
IE-rings vary only a few percent.

\begin{figure}
\centerline{\psfig{file=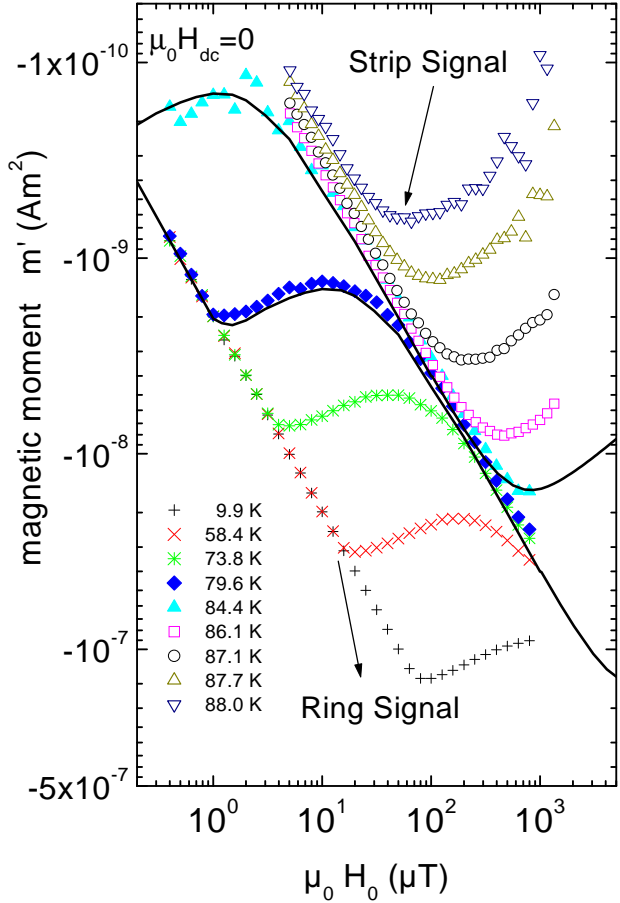,width=91mm}}
\caption{In-phase component of the magnetic moment of sample LA2 as a function of applied ac
  field at different temperatures. The continuos lines were calculated
  using Eq. (\protect\ref{eq:x}) assuming an overlap of the response of a ring
  with the corresponding strip signal 
\protect\cite{Brandt94apr1} with a ratio $J_c^{\rm strip} \approx 3000
  J_c^{\rm ring}$.}
\label{mLA2}
\end{figure}

On the origin of the remarkable difference in
$J_c$ of the rings structured by laser ablation we speculate as follows. During treatment by
the laser the tiny region
within the laser spot is heated up to several thousands  Kelvin for a few
milliseconds. Because of the bad thermal conductivity of the substrate
the temperature of the film rises to several hundred degrees in a
small region around the spot as well.
\begin{figure}
\centerline{\psfig{file=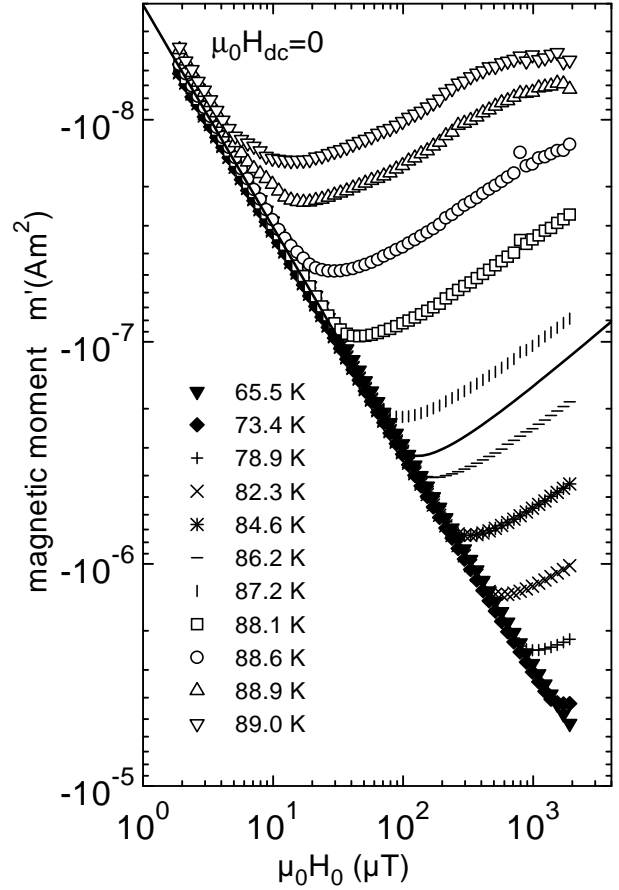,width=91mm}}
\caption{In-phase component of the magnetic moment of sample IE1 as a function of applied ac
  field at different temperatures. The continuos line was calculated
  using Eq. (\ref{eq:x}).} 
\label{mIE1}
\end{figure}
It is known that heating can change the oxygen
concentration of ${\rm YBa_2Cu_3O_{7-\delta}}$. As a result the
superconducting properties at the edges of the structured path may be altered up to the complete loss of
superconductivity. Also, thermally induced shock waves may produce microcracks within
the width of the ring and degrade the maximum critical current density.
Our measurements of  the thickness profile of the
ring after structuring indicate a region of $\rm \leq 5 \mu
m$ at the edges of the ring where the film appears to be completely damaged.
In contrast we found sharp edges on films structured by ion beam
etching. 

\begin{figure}
\centerline{\psfig{file=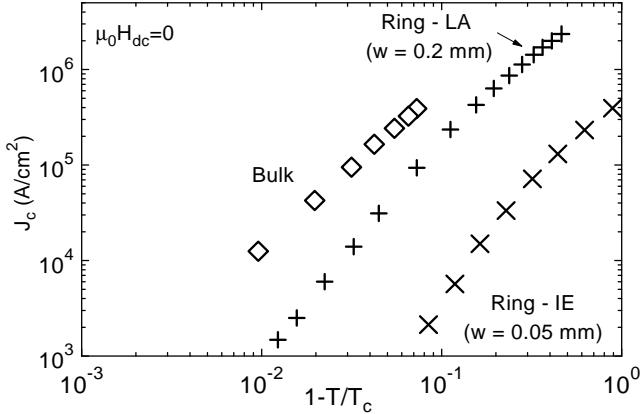,width=91mm}}
\caption{Comparison of the critical current density as a function of
  reduced temperature for consecutive patterning sample LA1 by
  laser ablation (Ring-LA) and ion beam etching (Ring-IE).} 
\label{jcLA1}
\end{figure}

To compare both structuring methods we can use a better
way to present the data by plotting the in phase
component of the magnetic moment $m' = \chi'H_aV$ as a function 
of the ac magnetic field amplitude.
This is shown for the rings~LA2, Fig.~\ref{mLA2} and~IE1,
Fig.~\ref{mIE1}.
The LA-ring in Fig.~\ref{mLA2} shows a clear two-step-transition,
one in the low-field-range and the other in the high-field range. We
recognize a crossover from the ring with a lower $J_c$ to the bulk
signal for a long strip \cite{Brandt94apr1}, see Fig.~\ref{mLA2}.  On the other hand the 
IE-ring in Fig.~\ref{mIE1} shows only a slight deviation from the expected ideal ring
behavior (continuous line in Fig.~\ref{mIE1}) very probaly due to small inhomogeneities within the
ring.

To check whether the  laser ablation structuring process leads to a
``homogeneous'' or ``inhomogeneous'' $J_c$ over the
width of the ring we structured sample LA1 once more by ion beam
etching from a width of 0.2 mm to 0.05 mm. Whereas 
in the ``homogeneous'' case 
we do not expect a change in the measured $J_c$ in the
``inhomogeneous'' case a 
different value for $J_c$ should be measured due to the existence of
islands with higher or lower critical current density. In Fig.~\ref{jcLA1} the results 
for the two consecutive structuring procedures
are shown. We assign the observed decrease in $J_c$ after the second
structuring to tiny islands of  lower $J_c$ which are along 
the path of the ring with smaller width and were created after the first laser ablation
structuring, indicating  strongly
inhomogeneous film properties.

\begin{figure}
\centerline{\psfig{file=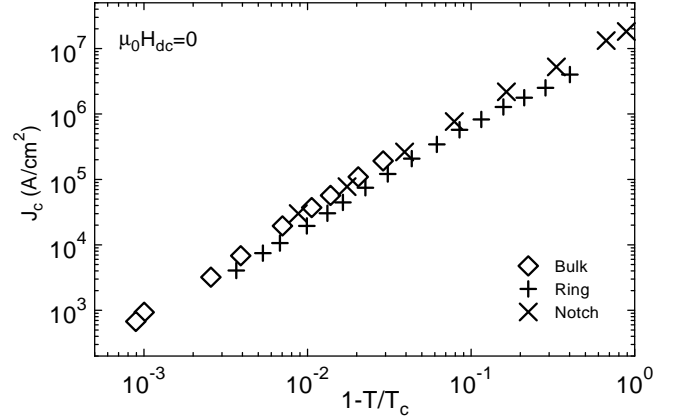,width=91mm}}
\caption{Comparison of the critical current density as a
  function of reduced temperature of the ring prepared by ion beam etching in
  which we subsequently cut a notch (sample N).}
\label{jcnotch}
\end{figure}

In a similar experiment we reduced the width of the IE-ring~(notch)
by cutting a notch into it by ion beam etching,  leaving only a small (16 $\mu$m)
path where the 
current can flow. By this means we basically measure $J_c$ in the path
region. 
Figure~\ref{jcnotch} presents the critical current density
corresponding to bulk, ring and ring-notch signal. Within experimental
uncertainty $J_c$ remains unchanged. This result supports the
assumption of a homogeneous $J_c$ in the film. 
Ion beam etching is therefore a far less destructive structuring
technique than laser ablation.  

\subsection{Field Dependence of the Pinning Potential}\label{spinning}

\begin{figure}
\centerline{\psfig{file=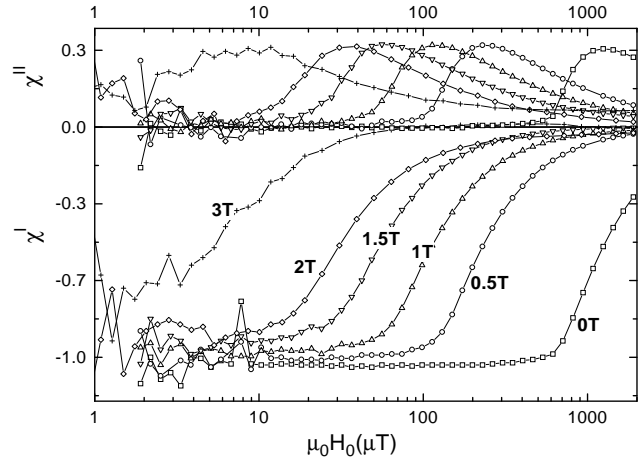,width=96mm}}
\caption{Ac field amplitude dependence of the normalized ac susceptibility
\mbox{$\chi=\chi'-i\chi''$} for dc fields up to 3T at constant temperature $T = 79.8$K, Ring IE2.}
\label{pdcshift}
\end{figure}

For the measurements presented in this section we have chosen  rings
with highest critical current density and with  different radius to width
ratios, e.g. rings~IE1 and~IE2. Together with a dc bias field the small
ac perturbation $\mu_0H_{0}<2$mT was applied. In Fig.~\ref{pdcshift} one recognizes that
the dc bias field mainly shifts ${\chi_{max}}''$ to lower ac amplitudes and
leaves the dependence of $\chi$ on the ac field amplitude essentially unaltered. This behavior
is due to a decrease in the critical current density that determines the field
$H_p \propto J_c$ at which the magnetic flux enters into the center of the ring.
 
During the increase of  the applied dc
field a current is induced in the ring. As described in section \ref{s:ac} perfect shielding occurs as
long as the current density is below the critical one. At
currents larger than $J_c$ flux can move into the center until the critical
value is reached again. At large dc fields the magnetic moment of the
ring thus saturates at a value that is determined by the shielding
capability of the ring which is given by the critical current related to the
specific dc field. By applying then a harmonic ac signal to the ring,
 its magnetic moment has a hysteresis loop around the static dc bias field similar as for $H_{dc}=0$.

\begin{figure}
\centerline{\psfig{file=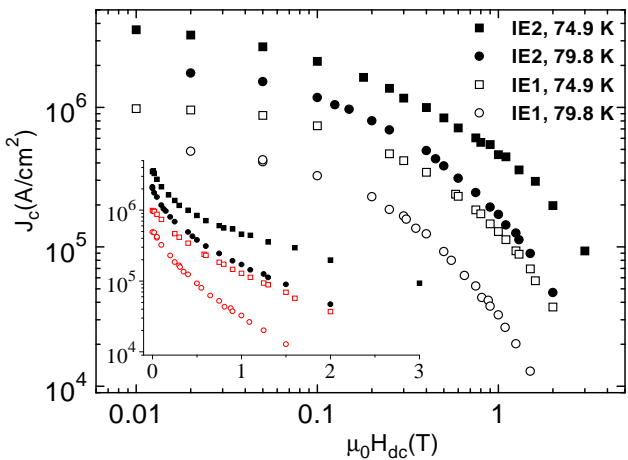,width=103mm}}
\caption{Dc field dependence of the critical current
density at two different temperatures in a log-log plot, derived from the shift of the
penetration field $H_p$ for the rings IE1 and IE2. The inset shows the same data
plotted in a semi-log plot.}
\label{pfieldjc}
\end{figure}

Since the application of the dc field leads to flux penetration it
will influence the pinning of vortices in the ring material as
well. By this means the flux creep exponent $n=U_0/kT$ is altered. In
Fig.~\ref{pdcshift} an increase in flux creep can be found for large
dc fields. As predicted by Brandt \cite{Brandt97jun1} a slight smoothing of
\mbox{$\chi''$} or $\chi'$ around the penetration field $H_p$ is observed for increasing dc field, i.e. 
 decreasing~$n$. From the
shift of $H_p$ with dc field one can determine the field dependence of
$J_c$ shown in
Fig.~\ref{pfieldjc}. The field dependence of $J_c$ we obtain is similar to that 
measured by vibrating reed experiments in Y123 thin films \cite{Ziese94nul2}.
The observed field dependence of $J_c$ can be understood within the collective
pinning theory for three dimensional pinning mechanism \cite{feige}. Three
dimensional pinning is assumed since the correlation length of the
vortex lines along the field direction is smaller than the thickness of the films \cite{Ziese94nul2}.
The magnetic field independence of $J_c$ observed at low fields indicate that
the vortices are pinned independently, i.e. the single vortex pinning regime. 
At higher fields the interaction between flux lines leads to the formation of
flux bundles which are pinned collectively and therefore $J_c$ decreases with
field.

\begin{figure}
\centerline{\psfig{file=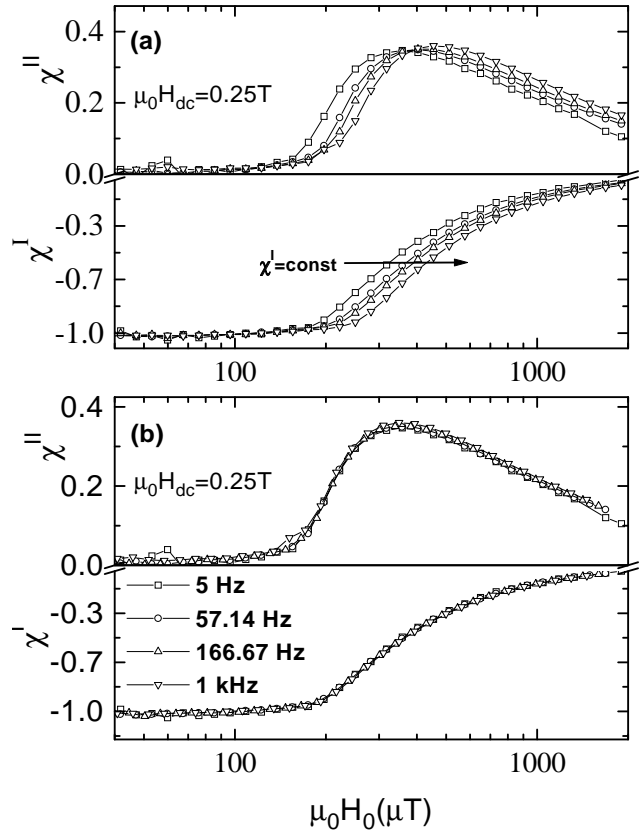,width=91mm}}
\caption{(a) Ac field amplitude dependence of the normalized ac susceptibility for
frequencies $5; 57.14; 166.67; 1000$~Hz for  the ring~IE2, at $T$=79.8K and  $\mu_0 H_{dc}=0.25$T.
 The value $\chi'$ at which the frequency
shift is analyzed is indicated by the horizontal arrow in the plot. The
corresponding ac field amplitudes are ${\mu}_0H_{0}=300; 338.9; 358.1; 395.8~
\rm \mu T$. (b) Re-scaled plots using Eq.~(\ref{esusc1}) with $n=20.2$ derived
as outlined in section \ref{stheory}.}
\label{psuscf}
\end{figure}

\begin{figure}
\centerline{\psfig{file=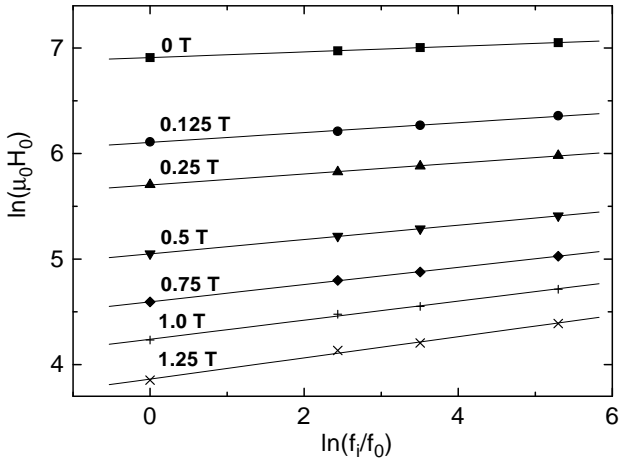,width=105mm}}
\caption{Frequency shift of the
ac field amplitude in a log-log scale for the ring~IE2 at $T$=79.8K. The frequencies are
$f_0=5$Hz, $f_i=57.14; 166.67; 1000$Hz. From the slope of the obtained
linear fit the flux creep exponent $n$ is determined, see section
\ref{stheory} for details. For $0.25$T the corresponding ac amplitude
scans are shown in Fig.~(\ref{psuscf}).}
\label{pfshift}
\end{figure}

In what follows  we concentrate on the field dependence of  the pinning potential $U_0$. For a series of dc
fields we measured the amplitude dependence of $\chi'$ at four
different frequencies. Such a scan can be seen in Fig.~\ref{psuscf}(a)
for the ring~IE2. In a step by step procedure we determine
the ac field amplitudes corresponding to the frequencies for a fixed value of
$\chi'$ where the slope is largest. This is indicated by the arrow  in
Fig.~\ref{psuscf}(a). Then we plot the logarithm of
the obtained magnetic ac field amplitudes versus the logarithm of the respective
frequencies, see Fig.~\ref{pfshift}, and by performing a linear fit we determine the
corresponding slope $\beta=1/(n-1)$ for different dc fields. 
Within the available frequency range this plot
gives straight lines for fixed dc field and temperature as
theoretically \cite{Brandt97jun1} expected. 
This gives the flux creep
exponent $n$ for fixed temperature and dc field, see Eq.~(\ref{hsca}). 
Figure~\ref{psuscf}(b)
shows the rescaling of all four curves in (a) following Eqs.~(\ref{ecvlaw}) and 
(\ref{etransform}) and using the obtained value
for $n$. 
In routine measurements
one only needs to measure $\chi$ around the maximum in $\chi''$ 
 to determine the shift of $\chi'$ with frequency (only 15 to 20 points per scan). The
necessary information on the ac field amplitude range to be measured at each dc field can be
obtained by a preliminary scan. Within 20 hours we were thus able to
measure the scans required to determine $n(T,H_{dc})$ for 10 dc fields
at one fixed temperature. 

\begin{figure}
\centerline{\psfig{file=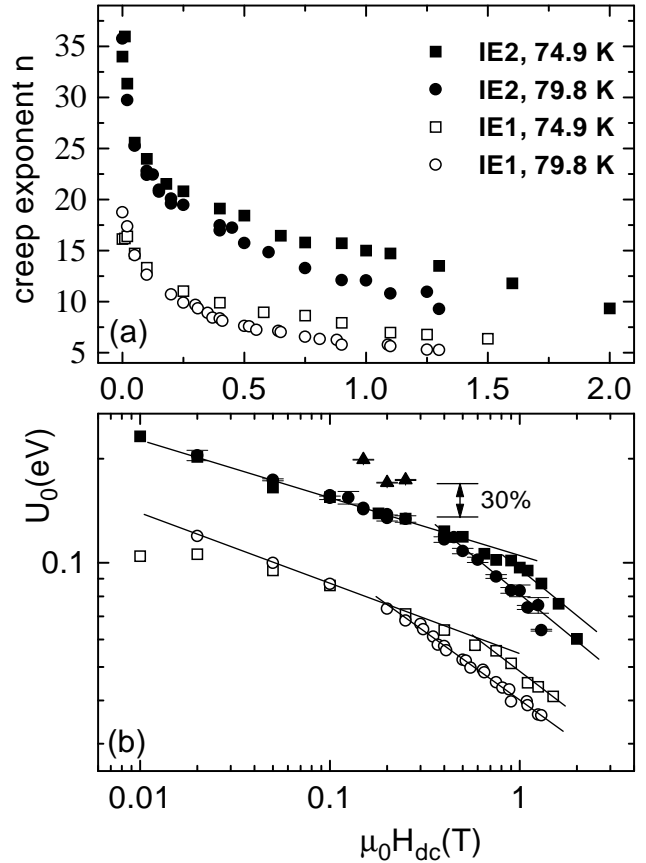,width=95mm}}
\caption{Dc field dependence of
(a) the flux creep exponent $n$ and (b) the pinning
potential $U_0$ for the rings~IE1 and IE2 at  $T$=74.9K and 79.8K. Values are 
derived from the scaling relation outlined in
section \ref{stheory}. Large $n$ at zero dc field correspond to a Bean-like 
behavior for the ac response of structured thin rings.  (b) Power law
fits (continuous lines) $U_0 \propto H_{dc}^{-\alpha}$ give $\alpha \approx 0.2$ and $0.4-0.7$
for the low- and high-field regions respectively. For
comparison the values $U_0(H_{dc})$ obtained from relaxation measurements, see
Fig.~\protect\ref{prelax}, 
are plotted  with solid up-triangles.}
\label{pufield}
\end{figure}

Scans for the two selected rings and temperatures \mbox{$T$=74.9K} and
\mbox{$T$=79.8K} have been performed and analyzed in this manner. The
dependence of the flux creep exponent $n(T,H_{dc})$ and the pinning potential
$U_0(T,H_{dc})$ on dc field are shown in Fig.~\ref{pufield}. In
Fig.~\ref{pufield}(a) one clearly sees the strong decrease of $n$ with
dc field. It can be also noted that $n$ increases the lower the temperature of the
ring. The large values of $n \approx 20-40$ at
$H_{dc}=0$ for both rings are consistent with the little frequency dependence of the ac
response at low dc fields. For this reason the Bean model is a good approximation in
this limit. At the same $T$ and $H_{dc}$ we find \mbox{$n_{IE2}>n_{IE1}$}. We assign this 
difference in $n$ between the two rings to
variations in the film properties from which the rings were
structured. 
\begin{figure}
\centerline{\psfig{file=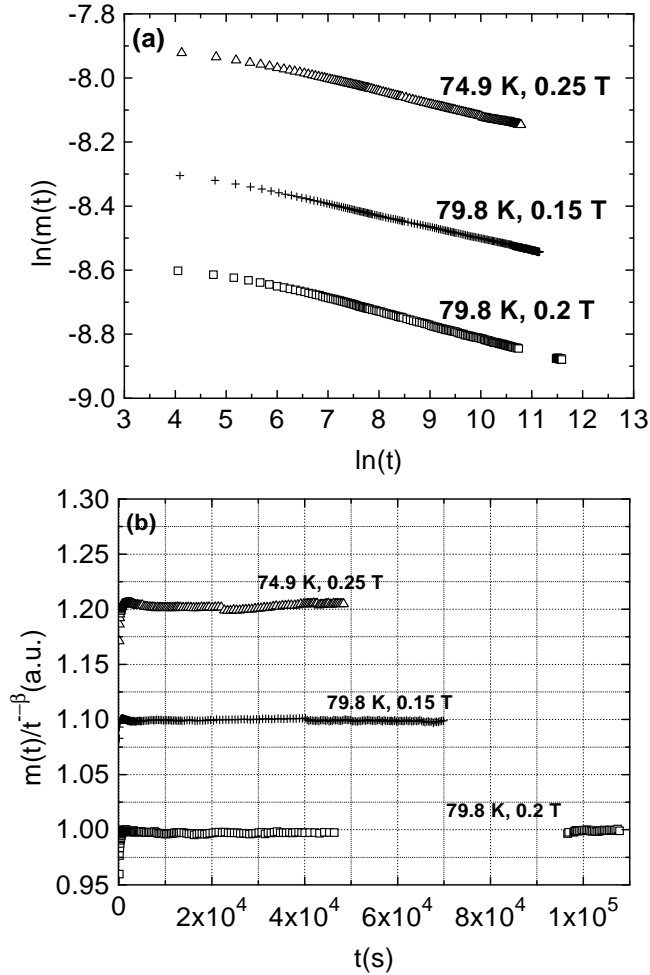,width=103mm}}
\caption{(a) Log-log plot of the relaxation of the magnetic moment $m(t)$ for the ring~IE2, after
the external dc field has been decreased to the constant value
$\mu_0H_{dc}$ as indicated in the figure. From the slope we obtain
$n = 26.9~(74.9~$K, 0.25~T),  28.9 (79.8~K, 0.15~T), 24.8 (79.8~K, 0.20~T), 
see Eq.~(\ref{erelax}). (b) Time independence of
$m(t)/t^{-\beta}$ calculated with the
respective exponents $\beta = 1/(n-1)$ obtained from (a). The curves were normalized for convenience.}
\label{prelax}
\end{figure}
The same qualitative statement holds for $J_c$ as well,
see Fig.~\ref{pfieldjc}. In Fig.~\ref{pufield}(b) one notes for both
rings a similar dependence of $U_0$ on $H_{dc}$. We 
distinguish a low- and a high-field region in which we obtain
straight lines in a double logarithmic plot. Assuming a power
law for the field dependence of $U_0$, e.g.~$U_0 \propto H_{dc}^{-\alpha}$, we obtain exponents 
$\alpha \approx 0.2$ and $\alpha \approx 0.4-0.7$ in the low- and
high-field regions, respectively. At $ T = 100~$K and in a dc field range between
20 and 100~Oe a similar dc field dependence for the flux creep exponent $n \propto H_{dc}^{-0.18}$
has been recently observed for HgBa$_2$CaCu$_2$O$_{6+\delta}$ 
thin films ($T_c = 120~$K) \cite{Jonsson99}.

Figure~\ref{pufield}(b) also shows
the values for $U_0$ obtained from relaxation measurements. We
used this technique to verify our results by an independent method. In the 
presence of flux creep the
magnetic moment $m(t)$ relaxes after ramping of the external field
$H_{dc}$ to some value. After a transient time $\tau$ one observes
the universal relaxation \cite{Brandt96jun,Brandt97jun1}
\begin{equation}\label{erelax}
m(t) \propto (t/\tau)^{-\beta}\,.
\end{equation}
With a SQUID magnetometer we have performed three relaxation measurements with the ring~IE2. The
results are shown in Fig.~\ref{prelax}(a). Following
Eq.~(\ref{erelax}) we obtain the flux creep exponent $n$ and hence the
pinning potential $U_0$ from a linear fit in the double logarithmic plot. 
With the respective values for
$n=(1+\beta)/\beta$, $m(t)/t^{-\beta}$ vs.~$t$   should be time independent. This is nicely 
confirmed in
Fig.~\ref{prelax}(b). A similar time dependence with $n \simeq 20$ at $H_{dc} = 1~$T 
has been obtained for the irreversible magnetization of Bi$_2$Sr$_2$Ca$_2$Cu$_3$O$_{10}$ 
samples below 20~K \cite{Makarov96}. As shown in Fig.~\ref{pufield}(b), the 
relaxation measurements provide values for the pinning potential $U_0$ which are similar
to those obtained with the nonlinear ac susceptibility scaling procedure.

\section{Conclusions}
In this work we have used the nonlinear ac susceptibility method on structured YBa$_2$Cu$_3$O$_{7-\delta}$ 
high-temperature superconducting rings to study  the influence of different patterning methods
to the critical current density of the films. The nonlinear susceptibility method applied on 
superconducting rings of small enough widths is a suitable method to identify regions with 
degraded superconducting properties. We have found that laser ablation structuring degrades the
superconducting properties of the film, decreasing strongly its critical current density and
introducing inhomogeneities within $\sim 100 \mu$m from the patterned edges. In contrast, the ion 
beam etching procedure is a far less destructive technique. Measuring the nonlinear susceptibility 
of the rings at different frequencies we reconfirmed the recently proposed scaling relation for the
frequency and field amplitude dependence. The scaling relation allows us to obtain the field
dependence of the flux creep exponent or pinning potential at different temperatures. Our results
show a power law dependence for the flux creep exponent $n \propto H_{dc}^{-0.2}$ for fields
below $\sim 0.2~$T and at reduced temperatures $0.8 < T/T_c < 0.9$.

\section*{Acknowledgments}
The authors thank M. Lorenz for providing us with the thin films. This work was supported
by the German-Israeli Foundation for Scientific Research and Development (Grant G-0553-191.14/97)
and by the Innovationskolleg "Ph\"anomene an den Miniaturisierungsgrenzen" (DFG IK 24/B1-1).

\onecolumn

\begin{table}
\caption{Characteristics of the measured rings. The structuring methods
  were laser ablation (LA) and ion beam etching (IE). $R$ denotes the
  radius of the ring, $w$ its width, $T_c$ and $J_c$ (77 K) the
  critical temperature and critical current density at zero applied
  field. $J_c$ was calculated using Eq.~(\ref{eq:jc}).}
\label{samples}
\begin{tabular}{cccccccc}
sample & structuring & $R$ & $w$ & \multicolumn{2}{c}{before
structuring} & \multicolumn{2}{c}{after structuring}\\
& method & (mm) & (mm) &  $T_c$(K) & $J_c$ (77 K) & $T_c$ (K) &
$J_c$(77 K)\\ 
LA1 & LA & 1 & 0.2 & $\approx$ 89& $4.5 \times 10^6 $A/cm$^2$ &
$\approx$ 89& $3.5 \times 10^5 $A/cm$^2$ \\
& IE & 1 & 0.05 & & & $\approx$ 89 & $1.0 \times 10^4 $A/cm$^2$\\
LA2 & LA & 1 & 0.1 & $\approx$ 89 & $3.3 \times 10^6 $A/cm$^2$ &
$\approx$ 89 & $7.9 \times 10^3 $A/cm$^2$ \\
IE1 & IE & 0.9 & 0.18 & 89.7 & $3.3 \times 10^6 $A/cm$^2$
& 89.8 & $1.0 \times 10^6 $A/cm$^2$ \\
IE2 & IE & 0.9 & 0.1 & 89.8 & $3.0 \times 10^6 $A/cm$^2$ & 89.9 &
$1.8 \times 10^6 $A/cm$^2$ \\
N & IE & 0.95 & 0.075 & $\approx$ 90 & $2.8 \times
10^6 $A/cm$^2$ & 89.9 & $1.1 \times 10^6 $A/cm$^2$ \\
& notch & 0.95 & 0.016 & & & 89.9 & $1.9 \times 10^6 $A/cm$^2$ \\
\end{tabular}
\end{table}

\end{document}